\documentclass[12pt,a4paper]{report}
\usepackage{a4wide}
 \usepackage[english]{babel}
\usepackage[pdftex]{graphicx}
\usepackage[hang,small,bf]{caption}
\usepackage{amsfonts}
\usepackage{amsmath,amsthm,amssymb}
\usepackage{fancyhdr}
\usepackage[breaklinks = true]{hyperref}
\usepackage{achemso}
\usepackage{natbib}
\usepackage{chapterbib}
\usepackage{wrapfig}

\fancypagestyle{plain}{%
\fancyhead{} 
}

\begin{document}
\vspace*{0.35in}

\begin{flushleft}
{\Large
\textbf\newline{Sulfo-SMCC Prevents Annealing of Taxol-Stabilized Microtubules \emph{In Vitro}}
}
\newline
\\
Meenakshi Prabhune, 
Kerstin von Roden,  
Florian Rehfeldt, 
Christoph F. Schmidt \textsuperscript{*}
\\
\bigskip
Third Institute of Physics - Biophysics, Georg August University, G\"ottingen, Germany

\bigskip

*christoph.schmidt@phys.uni-goettingen.de

\end{flushleft}

\section*{Abstract}
Microtubule structure and functions have been widely studied \textit{in vitro} and in cells. Research has shown that cysteines on tubulin play a crucial role in the polymerization of microtubules. Here, we show that blocking sulfhydryl groups of cysteines in taxol-stabilized polymerized microtubules with a commonly used chemical crosslinker prevents temporal end-to-end annealing of microtubules \textit{in vitro}. This can dramatically affect the length distribution of the microtubules. The crosslinker sulfosuccinimidyl 4-(N-maleimidomethyl)cyclohexane-1-carboxylate, sulfo-SMCC, consists of a maleimide and an N-hydroxysuccinimide ester group to bind to sulfhydryl groups and primary amines, respectively. Interestingly, addition of a maleimide dye alone does not show the same interference with annealing in stabilized microtubules. This study shows that the sulfhydryl groups of cysteines of tubulin that are vital for the polymerization are also important for the subsequent annealing of microtubules.


\section*{Introduction}
Microtubules (MTs) are cytoskeletal filaments that are vital for cell division and vesicle transport in cells~\cite{alberts2013essential}. MTs are also believed to play a role in cell mechanics via interconnections with actin filaments~\cite{huber2015cytoskeletal}. Experiments and simulations have shown that the embedding of microtubules in actin networks lowers Poisson's ratio of the composite, compared to a pure actin network~\cite{pelletier2009microrheology,das2011mechanics}. MTs are hollow cylinders with a diameter of 25 nm, polymerized from heterodimers of $\alpha$ and $\beta$ tubulin. Microtubules are polar and have a plus and a minus-end. The polarity of MTs is crucial for their biological function, making it possible for molecular motors to travel along MTs in a specific direction. Polymerization and depolymerization kinetics at the ends of MTs are dependent on the state of the nucleotide (GTP/GDP) bound to each tubulin monomer, and GTP hydrolysis, coupled to polymerization, makes MTs non-equilibrium polymers. Microtubules show, under certain circumstances, dynamic instability, i.e., switching between phases of growth and shrinkage~\cite{desai1997microtubule}. In cells, MTs undergo various post-translational modifications and are the substrate for a multitude of MT-binding proteins that control many functions~\cite{westermann2003post}. Because \textit{in vivo} experiments are difficult to interpret due to the complexity of cells, protocols for polymerizing microtubules (MTs) \textit{in vitro} have been developed providing a more controlled system for the study of MT structure and function~\cite{koshland1988polewards,hyman1990preparation,hyman1991preparation}.

Stabilization of microtubules with the small drug paclitaxel (taxol) or with the non-hydrolyzable GTP analogue, GMPCPP~\cite{schiff1979promotion}, makes it possible to perform experiments with dilute solutions or single microtubules. The minus end of tubulin can be blocked by N-ethyl maleimide (NEM). When polymerized from a mixture of NEM-labeled tubulin and regular tubulin, MTs, therefore, exhibit unilateral growth at their plus end~\cite{hyman1990preparation}. Although the mechanism of inhibition of minus end assembly by NEM is not exactly known, it is believed that binding of NEM to $\beta$ tubulin Cys$^{239}$ is responsible for	the minus end capping effect~\cite{phelps2000nem}. A further property of MTs that is relevant for \textit{in vitro} reconstitution experiments is end-to-end annealing. Fragmenting MTs in high-shear flow, e.g. by pressing the solution through a fine syringe needle, results in a short average length that one can see increasing again within a few hours~\cite{rothwell1986end}.

Biopolymer network	architecture depends strongly on polymer length, especially for sparse networks~\cite{broedersz2008nonlinear,sharma2013elastic,heidemann2015elasticity}. \textit{In vitro} studies of such networks have been widely used to provide a basis for the understanding of cytoskeletal mechanics
~\cite{mizuno2007nonequilibrium,mackintosh1995elasticity,ruddies1993viscoelastic}. While many experiments have been performed on actin networks, crosslinked with a variety of actin binding proteins (ABPs)~\cite{shin2004relating,gardel2006prestressed}, networks of crosslinked microtubules have not been much explored. This is most likely due to the fact that in cells MTs are either strongly bundled in a parallel manner such as in the axons and dendrites of neurons or in the meiotic and mitotic spindle~\cite{conde2009microtubule,yamada1971ultrastructure,glotzer20093ms}, or they are sparsely dispersed as in the cellular transport network spreading out from the MT-organizing center~\cite{alberts2013essential}. In most cells, though, MTs will intermingle with both intermediate filaments and with the actin network, and the different polymer systems may mechanically influence and regulate each other~\cite{fakhri2014high,brangwynne2008nonequilibrium,brangwynne2006microtubules, yue2014microtubules, ezratty2005microtubule, wehrle2003actin, palazzo2001mdia, robert2014microtubule}. Published work on microtubule networks has focused on the characterization of network	viscoelasticity of rigidly crosslinked MTs~\cite{lin2007viscoelastic,yang2013microrheology}. A more realistic representation of the cytoskeleton would be a network of rigid rods with compliant crosslinkers. Theoretical models and simulations of composite networks of stiff rods and flexible crosslinkers exist~\cite{broedersz2009effective,heidemann2015elasticity,sharma2013elastic}, but experimental approaches in such heterogeneous networks have been limited. A way to construct such heterogeneous networks would be to chemically crosslink MTs with compliant polymers such as intermediate filaments or DNA.

In the process of constructing such networks, we have here tested the effect of a commercial crosslinker, sulfo-SMCC, on the length distribution of taxol-stabilized microtubules. Sulfo-SMCC is a commonly used hetero-bifunctional crosslinker bearing N-hydroxysuccinimide (NHS) ester and maleimide groups to react with primary amines and sulfhydryl groups, respectively~\cite{dogan2015kinesin,erben2006single,mamedova2004substrate}. We demonstrate in time-dependent measurements and dual-color experiments that sulfo-SMCC inhibits the end-to-end annealing of stabilized MTs. Curiously, addition of a maleimide or an NHS ester group alone does not show an equivalent inhibition of annealing.

\section*{Methods}

\subsection*{Preparation of microtubules}

Labeled and unlabeled porcine tubulin powders were commercially obtained from Cytoskeleton, Inc., Denver, CO, USA. Sulfo-SMCC was purchased from Thermo Fisher Scientific, Waltham, MA, USA. MTs were polymerized to a final concentration of 2 mg/ml in 80 mM PIPES buffer (pH 6.8) containing 10 $\mu$M taxol, 2 mM MgCl$_2$, 0.5 mM EGTA, and 1 mM GTP. A mixture of rhodamine-labeled tubulin (Cytoskeleton, Inc., Denver, CO, USA) and unlabeled tubulin (1:5) was used in the time-dependent measurements. In these experiments, polymerized MTs were divided in 2 groups: control and sulfo-SMCC (250 $\mu$M) treated. Both samples were imaged after diluting 1:5 and incubating for 0 h (within 45 minutes after polymerization), 6 h, and 24 h. We performed 3 independent sets of experiments, and the total number of scored microtubules per time point are given in Table~\ref{Table1}.

 For dual-color experiments, red and green microtubules were polymerized as described above, using rhodamine-labeled tubulin and HiLyte 488-labeled tubulin, respectively. The red and green microtubules were mixed in a 1:1 ratio, and then divided into control and sulfo-SMCC (250 $\mu$M) treated groups. Both samples were imaged after 0 h and 6 h. Data were collected from 3 sets of experiments (see Table~\ref{Table2} for further details).

 Last, a control experiment was performed with MTs treated with 250 $\mu$M Alexa 488 NHS ester dye (Life Technologies GmbH, Darmstadt, Germany), and 250 $\mu$M Alexa maleimide 488 dye (Life Technologies GmbH, Darmstadt, Germany) for 24 h. Details regarding the mean lengths calculated from one set of experiments and the total number of scored microtubules are given in Table~\ref{Table3}.

\subsection*{Imaging and image analysis}

Positively charged silane-coated coverslips were used to attach the negatively charged MTs to the substrate. MTs were imaged using an oil immersion objective (EC Plan-Neofluar 100x/1.3, Carl Zeiss MicroImaging GmbH, Jena, Germany) on a standard fluorescence microscope (Axiovert 200, Carl Zeiss). MT solutions were diluted to different degrees so that the recorded images were not too crowded, showing too many overlaps of MTs that would complicate the automated filament recognition. Images were recorded using a digital CCD camera (CoolSnap ES, Roper Scientific, Martinsried, Germany), and image analysis for measuring lengths of short MTs was done in Fiji with a custom script involving standard procedures~\cite{schindelin2012fiji}. Image processing involved thresholding, followed by forming 'masks' after excluding microtubules at the edges. The masks were then run through custom software, Filament sensor~\cite{eltzner2015filament}, which outputs the lengths of MTs. In case of dual-colored MTs, composite images of red and green channels were analyzed in the software to measure lengths of annealed MTs. Exclusion of tubulin clusters and length measurement of looped MTs was done manually. Length histograms were normalized by the total number of MTs scored and then fitted by a single exponential, modified to account for underestimating short MTs due to the resolution limit (Fig.~\ref{fig1}). The calculated mean lengths were compared using a standard t-test.


\section*{Results}

\begin{figure}[p]
\includegraphics[width=0.75\textwidth]{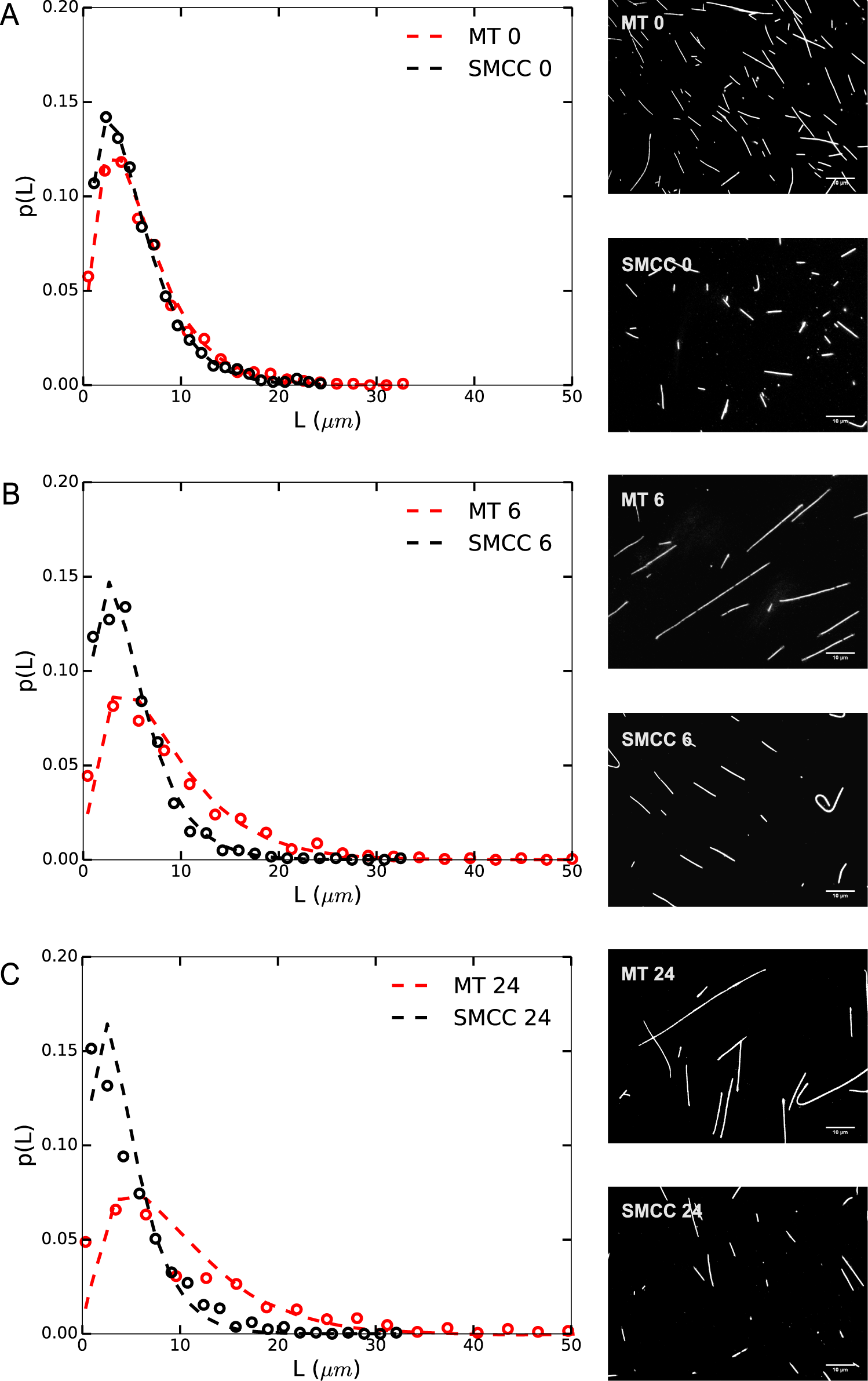}
\centering
\caption{{\bf Length distribution of sulfo-SMCC treated and untreated MTs after different incubation times.}
Length distributions of 2 mg/ml (tubulin) solution of (A) untreated and (B) sulfo-SMCC treated (250 $\mu$M) MTs after 0 h, 6 h, and 24 h are shown. Untreated MTs show a significant increase in their mean lengths after 6 h (p $<$ 0.001) and 24 h (p $<$ 0.001) of incubation, in comparison with those measured at 0 h. Sulfo-SMCC treated MTs show a constant mean length, independent of incubation time. Lengths of all samples are distributed exponentially; exponential fits with the normalized probability function $a^2Lexp(-aL)$ are shown as dashed lines. This functional form takes undersampling of short MTs due to the resolution limit into account.}
\label{fig1}
\end{figure}

MTs were treated with 250 $\mu$M sulfo-SMCC, and imaged after incubation for 0 h, 6 h, and 24 h. Fig.~\ref{fig1} shows the length distribution of sulfo-SMCC treated (A) and untreated MTs (B) after the different incubation times. Lengths of biopolymers that follow simple polymerization-depolymerization kinetics at fixed rates are distributed exponentially~\cite{oosawa1962theory,kawamura1970electron,kuhlman2005dynamic}. Consistent with previous studies~\cite{lin2007viscoelastic}, both treated and untreated MTs showed exponential length distributions at all time points. The mean lengths of untreated MTs increased significantly after 6 h (p $<$ 0.001) and 24 h (p $<$ 0.001) of incubation, in comparison with those measured at 0 h. Table~\ref{Table1} shows mean lengths and standard deviations of the data taken at each time point. The increase in length over time can be attributed to end-to-end annealing of MTs; a known phenomenon. Curiously, MTs treated with sulfo-SMCC showed a constant mean length, independent of the incubation time.

\begin{table}[h]
\centering
\begin{tabular}{|c|c|c|c|c|}
\hline
\textbf{Sample} & \textbf{Time (h)} & \textbf{Number of MTs} & \textbf{$L_{mean}$ ($\mu$m)} & \textbf{$L_{meanfit}$ ($\mu$m)}\\
\hline
MTs & 0 & 769 & 6.82 +/- 0.17 & 5.92 +/- 0.10\\
\hline
MTs & 6 & 880 & 9.77 +/- 0.24 & 8.21 +/- 0.34\\
\hline
MTs & 24 & 625 & 12.05 +/- 0.42 & 9.71 +/- 0.84\\
\hline
SMCC & 0 & 958 & 6.02 +/- 0.13 & 5.21 +/- 0.05\\
\hline
SMCC & 6 & 726 & 5.74 +/- 0.14 & 4.99 +/- 0.12\\
\hline
SMCC & 24 & 990 & 5.77 +/- 0.13 & 4.43 +/- 0.22\\
\hline
\end{tabular}\bigskip

\caption{\textbf{Lengths of untreated and sulfo-SMCC treated MTs at different time points.} The table shows the total number of microtubules (from three independent sets of experiments), average calculated mean length ($L_{mean}$) and average mean length from the fitting curve ($L_{meanfit}$) for sulfo-SMCC treated and untreated MTs, measured after 0 h, 6 h, and 24 h of incubation.} 
\label{Table1}
\end{table}

\begin{figure}[p]
\includegraphics[width=0.75\textwidth]{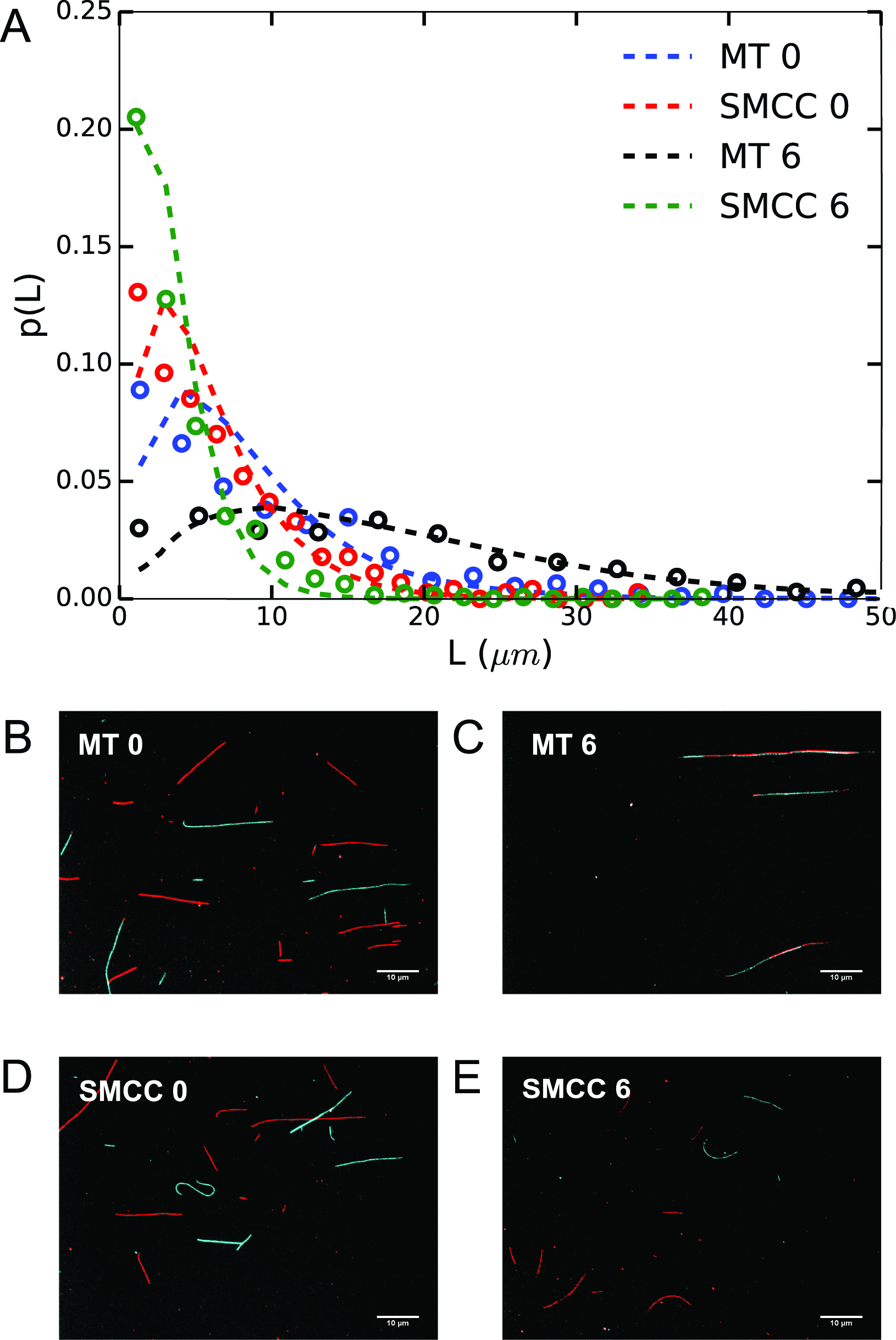}
\centering
\caption{{\bf Lengths of dual-colored MTs after 0 h and 6 h of incubation.}
(A) Length distributions of 2 mg/ml (tubulin) solution of untreated MTs and MTs treated with 250 $\mu$M sulfo-SMCC after 0 h and 6 h of incubation. Corresponding images of untreated MTs after 0 h (B) and 6 h (C). Sulfo-SMCC treated MTs after 0 h (D) and 6 h (E). Dotted lines are exponential fits to the data.}
\label{fig2}
\end{figure}

To verify that the observed effect was an inhibition of end-to-end annealing of MTs, we performed experiments using dual-colored MTs. A mixture of red and green microtubules, which were polymerized separately, was treated with sulfo-SMCC for 0 h and 6 h. The mean length of untreated MTs increased significantly after 6 h (p $<$ 0.001), as shown in Fig.~\ref{fig2}A. The average mean lengths of dual colored untreated and sulfo-SMCC treated MTs are given in Table~\ref{Table2}. Moreover, Fig.~\ref{fig2}C shows MTs consisting of both red and green (cyan in the images) segments after 6 h, clearly demonstrating the annealing process. Fig.~\ref{fig2}D shows that the mean length of sulfo-SMCC treated MTs did not increase after 6 h of incubation (p $>$ 0.05). Furthermore, the image of treated MTs after 6 h (Fig.~\ref{fig2}E) shows short filaments of either color, but not a mixture of both colors. These results strongly indicate that sulfo-SMCC hampers the temporal end-to-end annealing of microtubules.

\begin{table}[h]
\centering
\begin{tabular}{|c|c|c|c|c|}
\hline
\textbf{Sample} & \textbf{Time (h)} & \textbf{Number of MTs} & \textbf{$L_{mean}$ ($\mu$m)} & \textbf{$L_{meanfit}$ ($\mu$m)}\\
\hline
MTs & 0 & 337 & 10.62 +/- 0.46 & 8.28 +/- 0.68\\
\hline
MTs & 6 & 440 & 19.31 +/- 0.61 & 19.07 +/- 1.23\\
\hline
SMCC & 0 & 421 & 7.29 +/- 0.27 & 5.8 +/- 0.34\\
\hline
SMCC & 6 & 653 & 5.00 +/- 0.16 & 3.36 +/- 0.14\\
\hline
\end{tabular}\bigskip

\caption{\textbf{Dual-colored microtubule experiments.} The table shows the total number of microtubules (from three independent sets of experiments), average calculated mean length ($L_{mean}$) and average mean length from the fitting curve ($L_{meanfit}$) for different time points of sulfo-SMCC treated and untreated MTs.}
\label{Table2}
\end{table}

\begin{figure}[h]
\includegraphics[width=0.75\textwidth]{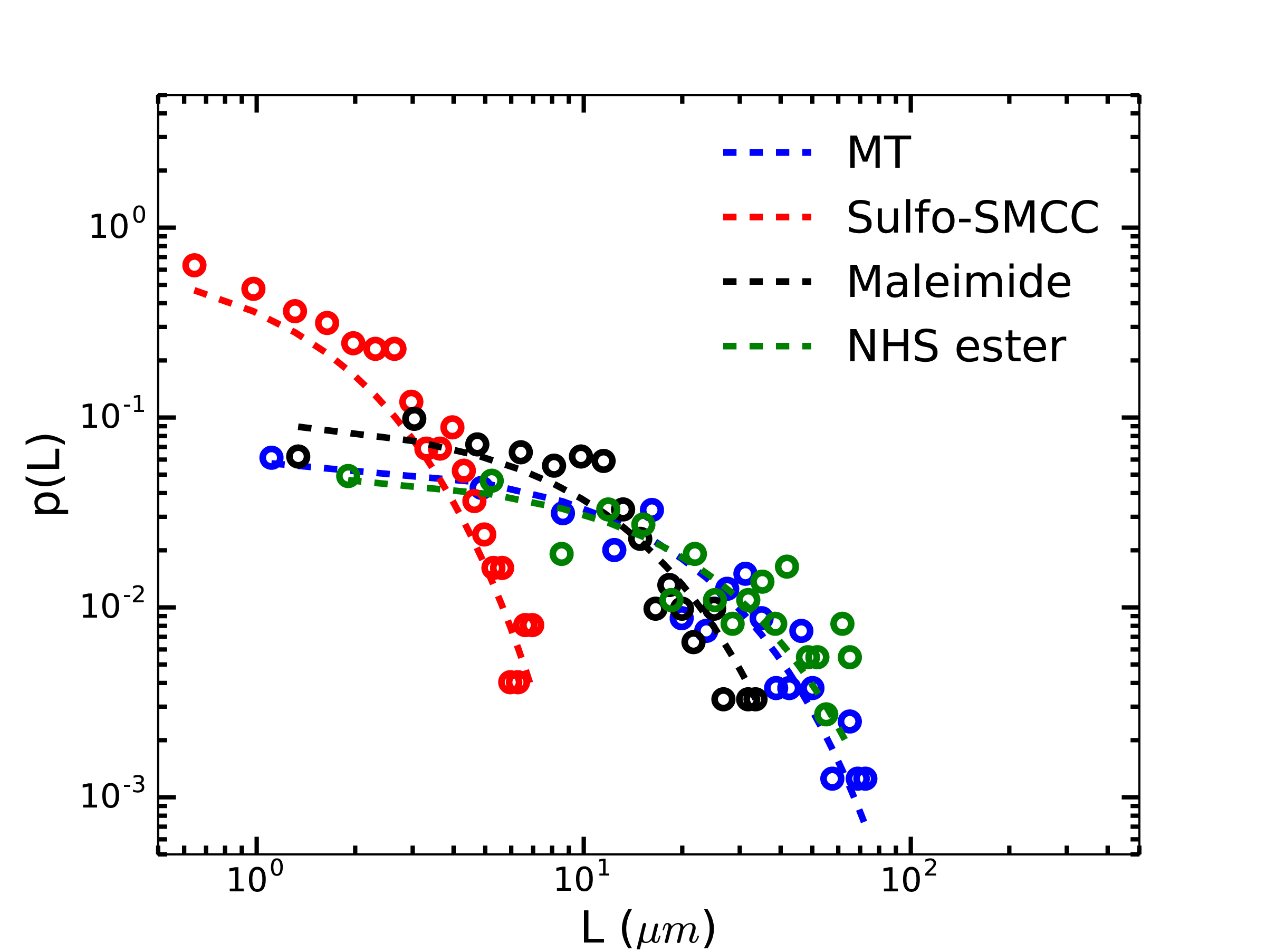}
\centering
\caption{{\bf Length distributions of chemically treated and untreated MTs after 24 h of incubation.}
Length distributions of 2 mg/ml (tubulin) solution of untreated MTs, MTs treated with 250 $\mu$M sulfo-SMCC, 250 $\mu$M maleimide dye and 250 $\mu$M NHS ester dye are shown. Lengths of MTs are distributed exponentially in all cases; single-exponential fits shown as dashed lines. Treatment with sulfo-SMCC resulted in drastically shorter MTs in comparison to untreated MTs (mean length = 2 $\mu$m and 17 $\mu$m, respectively). Maleimide dye treated MTs, however, showed a less drastic effect with an intermediate mean length of 9.2 $\mu$m. Addition of NHS ester dye to MTs as a control did not affect the lengths of MTs.}
\label{fig3}
\end{figure}

Lastly, we tested the effect of the individual reactive groups of sulfo-SMCC, i.e., maleimide and NHS ester on the length of MTs. MTs were incubated for 24 hrs with 250 $\mu$M of sulfo-SMCC, maleimide dye, or NHS ester dye. Control samples consisted of MTs without chemical treatment, also incubated for 24 h. Fig.~\ref{fig3} shows length distributions calculated from images of MTs under the three treatment conditions.

 We observed that MTs treated with sulfo-SMCC were drastically shorter compared to untreated MTs with almost an order of magnitude difference in their mean lengths (see Table~\ref{Table3}). Curiously, maleimide dye treatment at the same concentration resulted in only a two-fold shorter mean length than that of the untreated MTs. It is unclear why the effect of sulfo-SMCC is much more pronounced than that of maleimide dye although they bear the identical reactive group. We therefore tested a control sample to check for a possible effect of the NHS ester group of sulfo-SMCC on MT length. The length distribution after NHS ester treatment of MTs overnight, however, was very similar to that of untreated MTs. This excludes a possible effect of the NHS ester group.

\begin{table}[h]
\centering
\begin{tabular}{|c|c|c|c|c|}
\hline
\textbf{Sample} & \textbf{Time (h)} & \textbf{Number of MTs} & \textbf{$L_{mean}$ ($\mu$m)} & \textbf{$L_{meanfit}$ ($\mu$m)}\\
\hline
MTs & 24 & 213 & 17.01 +/- 1.04 & 16.3 +/- 1.12\\
\hline
SMCC & 24 & 747 & 1.99 +/- 0.04 & 1.3 +/- 0.24 \\
\hline
Maleimide dye & 24 & 210 & 9.2 +/- 0.46 & 9.75 +/- 1.18\\
\hline
NHS ester dye & 24 & 111 & 21.009 +/- 1.62 & 19.38 +/- 1.87\\
\hline
\end{tabular}\bigskip

\caption{\textbf{Chemically treated and untreated MTs.} The table shows the total numbers of microtubules (from one set of experiments), average calculated mean lengths ($L_{mean}$) and average mean lengths from the fitting curve ($L_{meanfit}$) for untreated MTs, and MTs treated with sulfo-SMCC, maleimide dye, and NHS ester dye for 24 h.}
\label{Table3}
\end{table}

\section*{Discussion}
We have shown through systematic experiments that sulfo-SMCC inhibits end-to-end annealing of taxol-stablized MTs. We speculate that the maleimide group on sulfo-SMCC binds to thiol groups on MT ends, and thereby prevents annealing. Previous studies have identified the sulfhydryl (SH) groups of cysteines of tubulin to be essential for MT polymerization~\cite{landino2007modification,huber20082,clark2014hypothiocyanous}. Chemically modifying cysteines on tubulin affected assembly of MTs \textit{in vitro} as well as \textit{in vivo}~\cite{kuriyama1974role,luduena1991tubulin}. Experiments with isothiocyanates have shown that mitotic arrest and apoptosis can be induced in cells by the covalent modification of cysteines in tubulin~\cite{mi2008covalent}. All prior experiments demonstrated deleterious effects of chemicals blocking the SH groups on MT polymerization. So far, little was reported regarding the effect on polymerized and stabilized MTs. Here, we have provided evidence that the vital SH groups on cysteine residues that are essential in MT polymerization appear to also be important for the annealing of stabilized MTs and that a commonly used crosslinker has a significant impact on MTs structure and dynamics. Therefore special care must be taken when using cysteine binding chemicals, not only during MT polymerization, but also after stabilization.

\section*{Acknowledgments}
We are grateful to Siddharth Deshpande, Abhinav Sharma, and Narain Karedla for advise on data analysis. This project was supported by the Deutsche Forschungsgemeinschaft (DFG) through Sonderforschungsbereich (SFB) 755 Project A3 and the Center for Nanoscale Microscopy and Molecular Physiology of the Brain (CNMPB).

\bibliography{references}

\end{document}